\documentclass[aps,prl,twocolumn,superscriptaddress,showpacs]{revtex4}
\usepackage{graphicx}
\usepackage{amsmath}
\usepackage{natbib}
\begin{document}
\title{Classical invariants and the quantization of chaotic systems}
\author{D. A. Wisniacki}
\affiliation{Departamento de Qu\'{\i}mica C--IX,
 Universidad Aut\'onoma de Madrid,
 Cantoblanco, 28049--Madrid (Spain).}
\affiliation{Departamento de F\'\i sica ``J. J. Giambiagi'',
 FCEN, UBA, Pabell\'on 1, Ciudad Universitaria,
 1428 Buenos Aires, Argentina.}
\author{E. Vergini}
\affiliation{Departamento de Qu\'{\i}mica C--IX,
 Universidad Aut\'onoma de Madrid,
 Cantoblanco, 28049--Madrid (Spain).}
\affiliation{Departamento de F\'{\i}sica,
 Comisi\'on Nacional de Energ\'{\i}a At\'omica.
 Av.\ del Libertador 8250, 1429 Buenos Aires (Argentina).}
\author{R. M. Benito}
\affiliation{Departamento de F\'{\i}sica,
 E.T.S.I.\ Agr\'onomos, Universidad Polit\'ecnica de Madrid,
 28040 Madrid (Spain).}
\author{F. Borondo}
\affiliation{Departamento de Qu\'{\i}mica C--IX,
 Universidad Aut\'onoma de Madrid,
 Cantoblanco, 28049--Madrid (Spain).}
\date{\today}
\begin{abstract}
Long periodic orbits constitute a serious drawback in Gutzwiller's
theory of chaotic systems, and then it would be desirable that other
classical invariants, not suffering from the same problem,
could be used in the quantization of such systems.
In this respect, we demonstrate how a suitable dynamical analysis
of chaotic quantum spectra unveils the fundamental role played by
classical invariant areas related to the stable and unstable
manifolds of short periodic orbits.
\end{abstract}
\maketitle

The correspondence between quantum and classical mechanics has been
a topic of much interest since the beginning of the quantum theory,
and more recently in relation to quantum chaos \cite{book1,book2}.
The question involves elucidating the classical objects and
properties on which to impose quantum restrictions,
this being at the heart of every semiclassical theory.

Very early, Ehrenfest noticed \cite{Ehrenfest} the importance of
classical adiabatic invariants, such as the action,
in the quantization of dynamical systems.
Later, Einstein \cite{Einstein} realized that the proper arena to
perform this quantization for integrable motions are invariant tori
\cite{LL}.
He also remarked that this theory is not applicable to chaotic motions,
due to the lack of supporting invariant classical structures.
After that, dynamical invariants are regarded as the geometrical
objects upon which reasonable semiclassical theories of quantum states
should be constructed.
Concerning the associated properties to be used, those which are
canonically invariant seem to be the natural choice, since they
render descriptions independent on the coordinate system.

Keeping within this scheme, Gutzwiller took in the 1970's a new
route, and chose periodic orbits (POs),
and their individual properties
(actions, Maslov indices and stability matrices),
leaving aside others (corresponding manifolds),
as the quantizable invariants \cite{Gutzwiller}.
In this way, he developed a semiclassical version of the quantum
mechanical Green function,
that has become the cornerstone of the semiclassical quantization of
chaotic systems.
The resulting density of states appears as the sum of contributions
of all POs of the system, and their repetitions.
The phase of each contribution is the action (symplectic area) along
the orbit (divided by $\hbar$), and the amplitude is proportional to
its stability.
Unfortunately, this theory suffers from a serious computational
problem in the exponential growth of the number of contributing
orbits with the Heisenberg time, $t_{\rm H}=2 \pi \hbar \rho (E)$,
with $\rho(E)$ the energy density.
This have precluded
its use except in very special situations \cite{Gutz2}.
In this respect, it is worth emphasizing that Gutzwiller's summation
formula can be used in two opposite ways.
In the direct route, it can be fed with classical information to
predict quantum eigenvalues.
Or alternatively, it can be used in an inverse way to extract
classical magnitudes from the eigenvalues spectrum \cite{Wintgen}.
Curiously enough, these two operations are not equivalent,
in the following sense.
When applied in the direct way, one needs to included longer and
longer POs (with periods of the order of $t_{\rm H}$) to
predict higher
energy eigenvalues.
However, when applied backwards, for example,
by Fourier transforming the eigenvalues spectrum of a chaotic billiard,
the periods (or other properties) of only short POs,
with values up to the magnitude of the Ehrenfest time, $t_{\rm E}$
\cite{tE}, are obtained \cite{Diego0}.
This asymmetry is not fully understood yet, and raises fundamental
questions about our present understanding of the quantum mechanics
of chaotic systems:
are long POs (with periods of the order of $t_{\rm H}$) relevant?,
or its inclusion in the theory of Gutzwiller is only a drawback?,
finally, responsible for the unreasonable computational effort
involved in the semiclassical computation of physical magnitudes.

In this Letter, we address this issue, by investigating the inverse
route in a non--standard way.
By explicitly including the dynamics of short POs (the only
relevant in this route) in the Fourier transform process,
we develop a method, relying only on purely quantum information,
able to extract the pertinent associated information
from the actual full quantum dynamics of very chaotic systems.
We have found conclusive evidence that the corresponding quantum
spectrum contains information about collective invariant objects
associated to short POs, namely, the homoclinic and heteroclinic
areas enclosed by their stable and unstable manifolds.
This implies some sort of interaction between periodic structures,
that can play a role equivalent to that of long POs in the Gutzwiller
formula.

To gauge the dynamical interaction between two POs, A and B,
in a quantum sense, we propose the use of the cross correlation
function
\begin{equation}
  C(t) = | \langle \phi_{\rm B}| \hat{U}(t) | \phi_{\rm A} \rangle |^2,
 \label{eq:CAB}
\end{equation}
where $\hat{U}(t)$ is the time propagation operator,
and $\phi_{\rm A,B}$ are suitable functions associated to the POs,
whose nature will be discussed later.
In our formula, the second part of the bracket follows the evolution
of the non--stationary function associated to one of the POs,
and the application of the bra extracts the information relative
to the other PO contained in it,
thus filtering out (at least to some extend) the rest.
By choosing a correlation function as our indicator, we have the same
information as in the corresponding spectra, but recast with a more
dynamical perspective.

A natural choice for $\phi_{\rm A}$ and $\phi_{\rm B}$ are
wavefunctions living in the vicinity of the corresponding POs.
These functions are constructed very efficiently,
either by dynamically averaging over the short time dynamics of the
associated PO \cite{Polavieja}, or by minimizing the energy dispersion
in a suitable basis of transversally excited resonances \cite{Vergini2}.
In this work, we use scar functions as defined in
Ref.~\onlinecite{Vergini2}.
These functions are highly localized in energy around some mean
values satisfying a Bohr--Sommerfeld type quantization rule
\begin{equation}
  \frac{S(E)}{\hbar} - \nu \frac{\pi}{2} = 2 \pi n,
\label{eq:2}
\end{equation}
where $S(E)$ is the dynamical action at energy $E$, $\nu$ the Maslov
index, and $n$ an integer number.

\begin{figure}[t]
  \includegraphics[width=7.5cm]{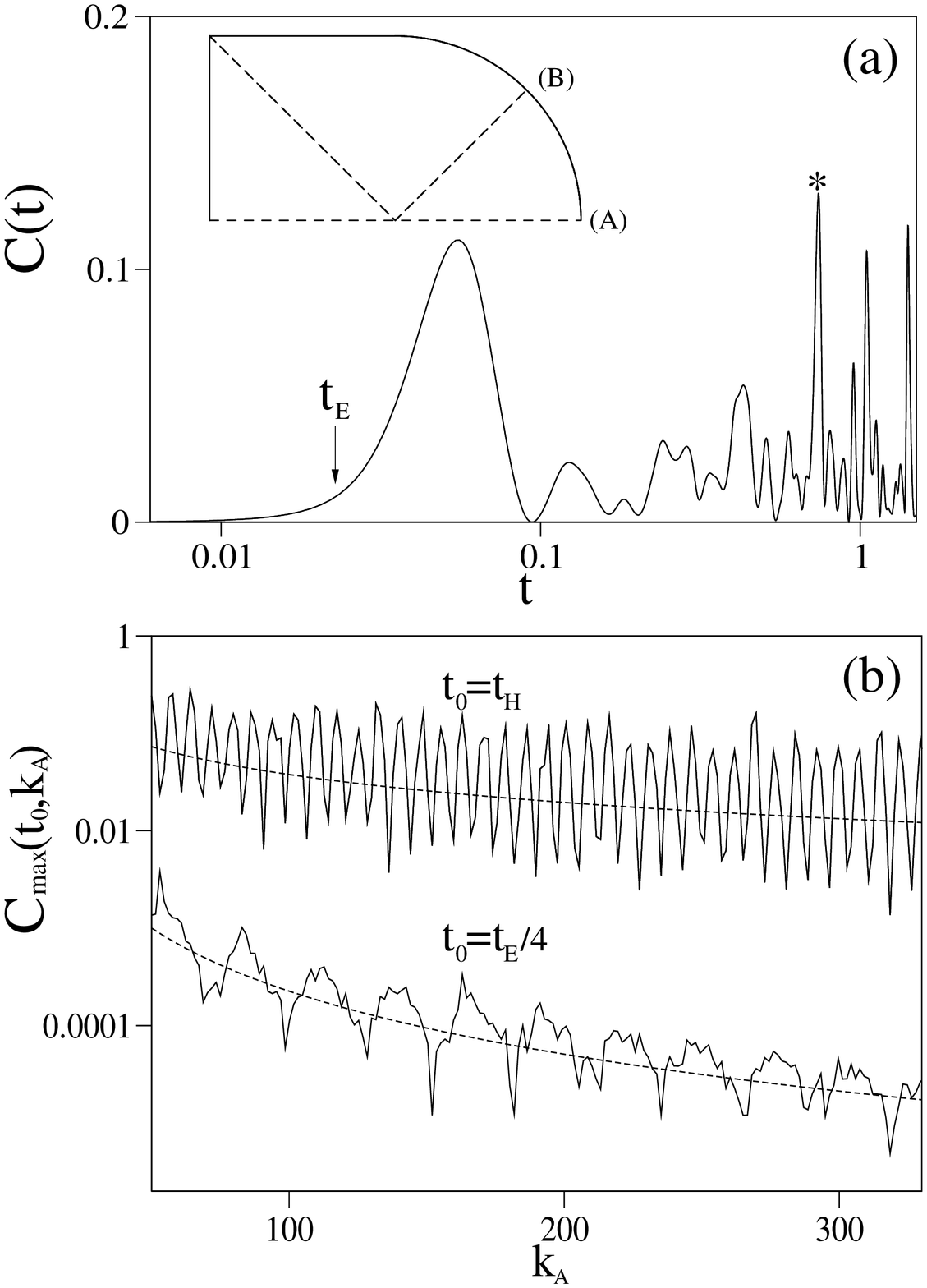}
 \caption{(a) Cross correlation function between the horizontal (A)
  and the V--shaped (B) periodic orbits shown in the inset,
  for a value of the wavenumber of $k_{\rm A}=155.116$.
  The time $t$ is measure in units of the Heisenberg time. \protect \\
  (b) Maximum value of the cross correlation function in the interval
  $[0,t_0]$ as a function of the wavenumber,
  for $t_0=t_{\rm E}/4]$ (lower curve) and $t_0=t_{\rm H}$ (upper curve).
  The mean decreasing tendency is indicated in dotted line.}
 \label{fig:1}
\end{figure}

In order to study the previous ideas, we choose a particle of mass 1/2
enclosed in a fully chaotic desymmetrized stadium billiard of radius
$r=1$ and area $1+\pi/4$, with Dirichlet conditions on the stadium
boundary and Neumann conditions on the horizontal and vertical
symmetry axis [see the inset in Fig.~\ref{fig:1} (a)].
For this system, the action takes a simple semiclassical relation
in terms of the mean wavenumber, $k$, and the length, $L$, of the
PO; namely, $S(E)/\hbar=k L$.

In our numerical study, we consider the horizontal (A) and V--shaped
(B) POs with lengths $L_{\rm A}=4$ and $L_{\rm B}=2 (1+\sqrt{2})$,
respectively [see the inset in Fig.~\ref{fig:1} (a)].
Let $\phi_{\rm A}$ be the scar function for $A$ with mean wave number
$k_{\rm A}$ [obtained from Eq.~(\ref{eq:2})], and $\phi_{\rm B}$ the
corresponding for $B$, with the mean wave number $k_{\rm B}$ closest
to $k_{\rm A}$.
We focus in the cross correlation function as defined in
Eq.~(\ref{eq:CAB}), and accordingly, we present in Fig.~\ref{fig:1} (a)
$C(t)$ for $k_{\rm A}=155.116$.
%
The most striking feature in the plot is the totally different behavior
exhibited by the correlation, below and above times of the order of
the Ehrenfest time, $t_{\rm E}$, (the actual value of which has been
marked with an arrow in the figure).
For short times, the correlation function increases monotonically from
zero up to the first maximum.
This maximum appears at approximately twice the value of $t_{\rm E}$,
point at which interference starts to be relevant.
After that, other maxima appear, and the behavior of the correlation
gets very complex for times of the order of the Heisenberg time,
$t_{\rm H}$, which is equal to one in the units system used by us.

\begin{figure}[t]
\includegraphics[width=7.5cm]{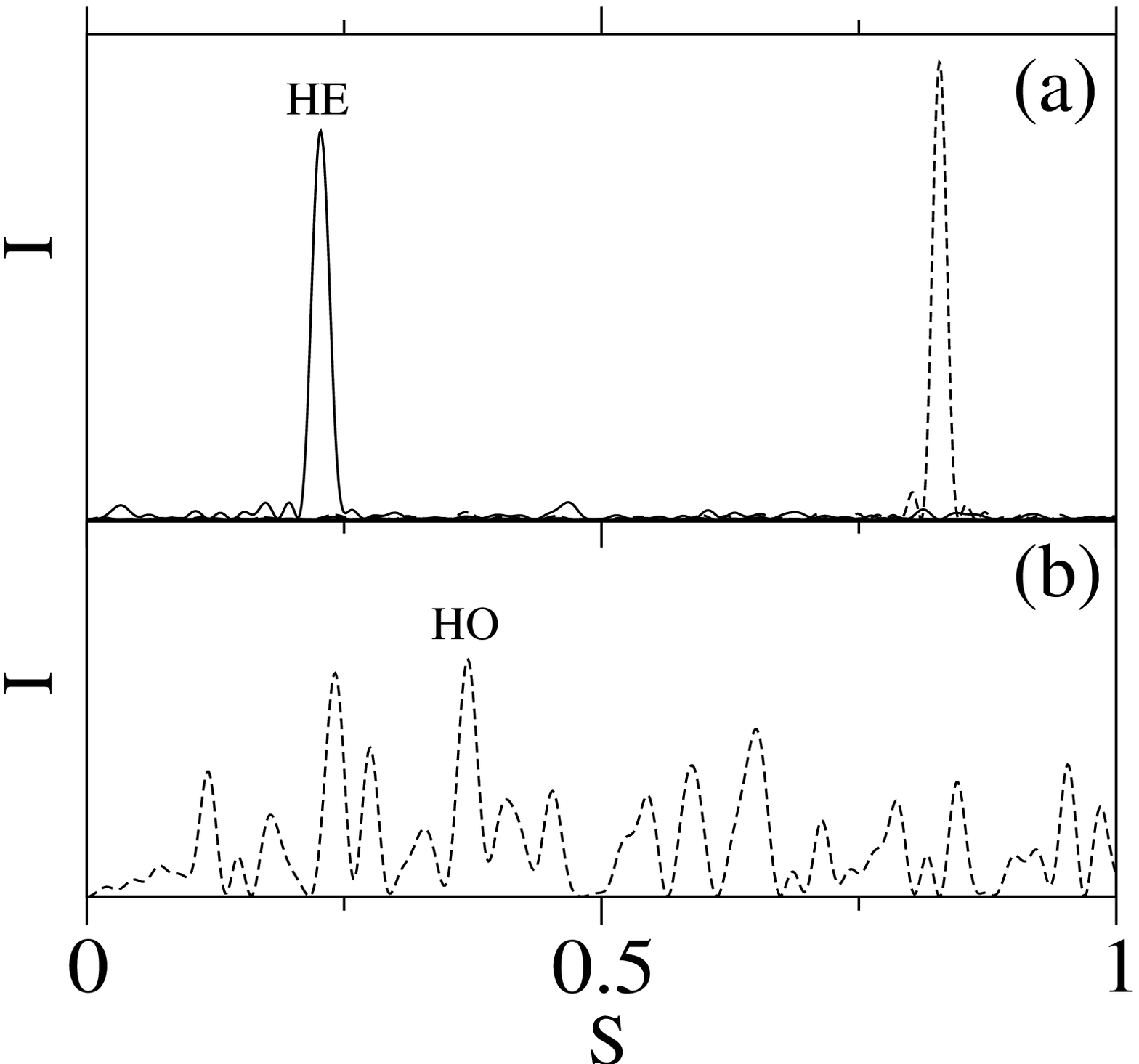}
 \caption{(a) Fourier transform of the maximum cross correlation
  functions in Fig.~\ref{fig:1} (b): $t_0=t_{\rm E}/4$ (full line),
  and $t_0=t_{\rm H}$ (dashed line).
    \protect \\
  (b) Rescaled intensity of the Fourier transform for $t_0=t_H$ in
   part (a) after the big peak has been removed.}
 \label{fig:2}
\end{figure}

\begin{figure}
  \includegraphics[width=5.0cm]{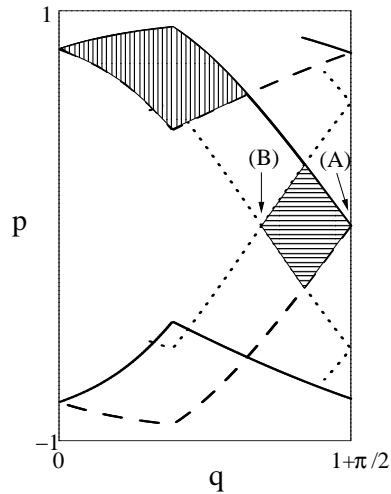}
 \caption{Phase space portrait in Birkhoff coordinates of the
  classical structures relevant to our calculations.
  The fixed points marked with (A) and (B) correspond to the labeled
  turning points of periodic orbits A and B shown in the inset to
  Fig.~\protect\ref{fig:1} (a).
  The unstable and stable manifolds of the horizontal PO are
  represented in full and dashed line, respectively, and those for
  the ``V''--shaped PO in dotted line.
  The shaded regions corresponds to the heteroclinic (horizontal lines)
  and homoclinic (vertical lines) referenced in the text.}
 \label{fig:3}
\end{figure}

To further characterize the interaction between POs,
some representative dynamically meaningful magnitude along the spectra
should be defined.
For this purpose, we take the maximum of $C(t)$ in the time interval
$[0,t_0]$
\[
  C_{{\rm max}}(t_0,k_{\rm A}) \equiv \max \{ C(t);
    \; {\rm for} \; 0<t<t_0 \},
\]
where the dependence on $k_{\rm A}$ has been explicitly included.
[For instance, in the case of $t_0=t_{\rm H}$, this maximum appears
marked with an asterisk in Fig.~\ref{fig:1} (a)].
Obviously, other quantities, such as for example the integral of
$C(t)$ in the interval, can be used as the representative magnitude;
our conclusions, however, are independent of this choice.
In Fig.~\ref{fig:1} (b) we show $C_{\rm max}(t_0,k_{\rm A})$,
as a function of $k_{\rm A}$, for two values of the maximum time $t_0$,
that have been taken equal to $t_{\rm E}/4$ (lower curve) and
$t_{\rm H}$ (upper curve).
As can be seen, both functions decay as $k_{\rm A}$ increases,
while oscillating at the same time with a dominant frequency.
To analyze the frequency dependence of these functions,
we have Fourier transformed them, after the signal has been properly
prepared by eliminating the decaying tendency (dotted line).
The resulting spectra are shown in Fig.~\ref{fig:2} (a) in full and
dashed line, respectively.
%
As can be seen, they both appears dominated by only one peak,
at values of the action:
$S=0.227$ for $t_0=t_{\rm E}/4$, and $S=0.828$ for $t_0=t_{\rm H}$.
Notice that $S$ has units of length due to the fact that the total
linear momentum of the particle has been set equal to one.
From the discussion above, our aim is to correlate these two peaks
with invariant classical structures related to the chosen POs.
Taking into account the numerical values of their positions,
the first peak (labelled HE in the figure) can be assigned to the
heteroclinic area, $S_{\rm AB}$, enclosed by the stable and unstable
manifolds emanating from the fixed points associated to POs A and B.
This region is shown, shaded with horizontal lines, in the phase space
portrait of Fig.~\ref{fig:3}, which illustrates the classical
structures relevant to our work.
%
When calculated, the corresponding area is 0.22540, value that agrees
extremely well with that numerically found for $S_{\rm HE}$.
Moreover, according to the results reported in
Ref.~\onlinecite{Vergini3} for a more abstract case,
this heteroclinic area is related to the semiclassical Hamiltonian
matrix element between scar functions through the relation
\begin{equation}
  | \langle \phi_{\rm B} | \hat{H} | \phi_{\rm A} \rangle |^2
     \propto \cos (S_{\rm AB} \; k),
 \label{eq:3}
\end{equation}
which presents the same oscillating frequency found by us.
This fact can be taken as a further confirmation of our previous
assignment.

Using the same kind of arguments as before, the peak on the dashed
curve of Fig.~\ref{fig:2} (a), corresponding to $t_0=t_{\rm H}$,
can be assigned to the difference in length between POs A and B.
This quantity amounts to $0.828426\ldots$, again in excellent agreement
with the value found numerically.
The existence of this peak, which is associated to the difference
in actions between orbits A and B [see Eq.~(\ref{eq:2})],
reflects the strong dependence of the interaction with
$k_{\rm A}-k_{\rm B}$ (which can also be related with the
energy separation between the resonant levels corresponding
to $\phi_{\rm A,B}$).
This effect has been further confirmed by analyzing the oscillatory
behaviour of this wavenumber difference, which turns out to be the
same exhibited by $C_{\rm max} (t_{\rm H},k_{\rm A})$,
although these two functions appear with opposite phases.

At this point it is worth to emphasize that the different origin of
the two peaks reflects the existence of two regimes in the cross
correlation function (\ref{eq:CAB}), with the corresponding transition
taking place at $t \sim t_{\rm E}$.
These two regimes can be easily understood in terms of the
Fermi golden rule, since $C_{\rm max}(t_0,k_{\rm A})$ is in fact a
measure of the probability transition between the resonant states
$\phi_{\rm A}$ and $\phi_{\rm B}$.
Accordingly, the behavior of $C_{\rm max}$ is given by the
competition between two factors:
the square of the coupling matrix element and the separation in energy
of the corresponding levels.

Let us now consider other components in the spectra of
Fig.~\ref{fig:2} (a). To observe them more clearly, we calculate
the rescaled intensity that is obtained after the biggest peak has
been removed.
When this is done for the two plotted curves only the results
corresponding to the dashed one are stable against local
variations of $t_0$.
They are shown in the lower part of the figure.
As can be seen, many different contributions appear, all with a
comparable order of magnitude.
Among them we have focussed in the highest one (labelled HO),
as the most interesting.
When the value of $S_{\rm HO}$ is compared to the relevant classical
invariants of our problem (see Fig.~\ref{fig:3}), we find that it
matches remarkably well with the homoclinic area enclosed by the
stable and unstable manifolds emanating from A;
this region appears shadowed with vertical lines in Fig.~\ref{fig:3}.
The assignment is also supported by the work by Ozorio de Almeida
\cite{Ozo}, which has shown how homoclinic motions can be quantized,
using the corresponding invariant manifolds.
We believe that all (or most) remaining peaks in the spectrum presented
in the lower part of Fig.~\ref{fig:2} can be interpreted in the same way,
using different homoclinic and heteroclinic regions corresponding
to the same POs.
Actually, we have succeded in assigning the two peaks located to the
left of that at $S_{\rm HO}$; however, a full description of the
procedure is deferred to a future publication.

In summary, some aspects concerning the role of short POs
in Gutzwiller's summation formula, the cornerstone for the
semiclassical quantization of chaotic systems, have been analyzed.
For this purpose, only purely quantum information has been used,
in order to obtain relevant classical invariants for a semiclassical
theory of quantum chaos.
We have found evidence that the short POs,
and their associated heteroclinic and homoclinic intersecting areas,
are relevant contributions to the spectra.
Furthermore, our results provide a semiclassical interpretation
on how structures localized along POs interact \cite{Diego2}.
This interaction is found to be given as the competition of two effects:
the energy separation between scar functions,
and the squared coupling matrix element between them,
following the celebrated Fermi golden rule.
Also, we have shown how the values of the corresponding parameters can
be theoretically evaluated {\it a priori}.
Finally, our results point out to the possibility of constructing
computationally tractable alternatives to Gutzwiller's theory,
in which the long POs are substituted by the interaction between
short POs.
In this respect, the steps given in Ref.~\onlinecite{Carlo} provide
a suitable frame in this direction that, together with a deeper
understanding of the interaction between POs, can be a bridge towards
a fully satisfactory semiclassical theory of chaotic systems based
on short POs.
This theory would be interesting not only from a fundamental point
of view, but also for its application, for example,
to nanotechnology \cite{book2}, where it has been recently shown
useful, for example, in the study of electron transport in
mesoscopic devices \cite{nano}.
\begin{acknowledgments}

This work was supported by MCyT and MCED (Spain) under contracts
BMF2000--437, BQU2003--8212, SAB2000--340 and SAB2002--22.
\end{acknowledgments}
%

%

\begin{thebibliography}
\bibitem{tE} The Ehrenfest time is defined as
$t_{\rm E} = \ln \hbar / \lambda$, being $\lambda$ the Lyapunov
exponent, and constitutes a rough indicator of the amount of time
needed by a wave packet to fill the available configuration space.
\end{thebibliography}
\end{document}